\documentclass[twocolumn,showpacs,prb,preprintnumbers,amsmath,amssymb]{revtex4-1}
\usepackage{graphicx}
\usepackage{dcolumn}
\usepackage{bm}
\usepackage{xcolor}
\usepackage{slashed}
\usepackage[mathscr,mathcal]{euscript}
\newcommand{\dslash}{\slashed{\partial}}

\def\sgn{\operatorname{sgn}}
\def\tr{\operatorname{tr}}
\allowdisplaybreaks{}
\begin{document}
\title{Hall effect, edge states and Haldane exclusion statistics in 
two-dimensional space}
\author{F. Ye}
\affiliation{Department of Physics, South University of Science and
  Technology of China, Shenzhen 518055,
  China}
\author{P. A. Marchetti}
\affiliation{Dipartimento di Fisica, INFN, I-35131 Padova, Italy}
\author{Z. B. Su}
\affiliation{Institute of Theoretical Physics, Chinese Academy of
  Sciences, 100190 Beijing, China}
\author{L. Yu}
\affiliation{Institute of Physics, Chinese Academy of Sciences, and Beijing National Laboratory for Condensed Matter Physics, 100190
 Beijing, China}
\affiliation{Collaborative Innovation Center of Quantum Matter, 100190 Beijing, China}

\begin{abstract}
  We clarify the relation between two kinds of statistics for particle
  excitations in planar systems: the braid statistics of anyons and the
  Haldane exclusion statistics(HES). It is shown non-perturbatively that
  the HES exists for incompressible anyon liquid in the presence of a
  Hall response. We also study the statistical properties of a specific
  quantum anomalous Hall model with Chern-Simons term by perturbation in
  both compressible and incompressible regimes, where the crucial role
  of edge states to the HES is shown.
\end{abstract}
\maketitle

\section{Introduction}  

Anyons are particle excitations obeying the \emph{braid
  statistics} which interpolates continuously between fermion and boson
statistics in two
dimensions(2D)\cite{Leinaas1977,Wilczek1982,Wu1984a,[{For a review see
    e.g.,}] Marchetti2010}. Abelian braid statistics can be
characterized by the phase factor $e^{i(\alpha-1)\pi}$ of the many-body
wave-function when two anyons are exchanged, with "any" $\alpha \in
[0,2)$, thus explaining their name. Fermions(bosons) correspond to
$\alpha=0$(1).  In real world anyons seem to appear in the fractional
quantum Hall systems\cite{Laughlin1983,Arovas1984}, and presumably on
the magnetized surface of topological insulators\cite{Qi2009}. Their
relation to high temperature superconductivity initiated decades
ago\cite{Laughlin1988,Wilczek1990}, is still being actively pursued
nowadays in a renewed form
\cite{Frohlich1992,Marchetti1998,Marchetti2011}, requiring the knowledge
of anyon thermodynamics, which however is barely known even for free
anyons with $0<\alpha<1$.

Another interpolation between fermion and boson statistics was proposed
by Haldane. It is known as \emph{Haldane's exclusion
  statistics}(HES)\cite{Haldane1991a} or Haldane-Wu
statistics(HWS)\cite{Wu1994}, based upon the following
state-counting ideas: For a many-body system in finite
volume, the dimension $d(N)$ of the Hilbert space of the $(N+1)$-th
particle depends on $N$, which leads to the definition of statistical
interaction $g$ via $\Delta d=(g-1) \Delta N$\cite{Haldane1991a}.
Obviously $g=0$ for fermions, $g=1$ for bosons, and other intermediate
values of $g$ define fractional HES.  Unlike the aforementioned braid
statistics, the HES is not limited to planar systems and one can
calculate their thermodynamic properties explicitly.  The representative
quantity is the energy distribution function $n_H(\epsilon,g)$ which was
derived by Wu\cite{Wu1994}. It turns out that particles with
HES(non-mutual) have a well-defined Fermi energy
$\epsilon_F$ at zero temperature if $g \neq 1$. When
$\epsilon<\epsilon_F$, $n_H(\epsilon,g)=1/(1-g)$, otherwise
$n_H(\epsilon,g)=0$, thus interpolating between Fermi-Dirac(FD)
distribution with $g=0$ and Bose-Einstein(BE) distribution with $g=1$ at
zero temperature.  At finite temperatures $n_H$ is much more complicated
than FD and BE distribution. In analogy with Landau Fermi liquid theory,
a theory of interacting HES particles called Haldane liquid was
developed in Ref. \onlinecite{Iguchi1998}. An important result is the
generalization of Luttinger theorem: the interaction does not change the
volume enclosed by the Fermi surface for HES.

In one dimensional(1D) systems a deep connection between Luttinger
liquids, braid statistics and HES was established in Refs.
\cite{WuYu1995,Wu2001}, but for planar systems, braid statistics and HES
are not always equivalent. In fact the free non-relativistic anyons do
not show the evidence of HES. On the other hand, the non-relativistic
anyon model in the strong magnetic field is exactly solvable after
projection to the lowest Landau
level(LLL)\cite{Johnson1990,Dunne1991,deVeigy1994} and it is described
by an equation of state consistent with HES\cite{deVeigy1994}. It was
shown that the filling factor of the LLL at $T=0$ equals to
$1/(1-\alpha)$\cite{deVeigy1994,Ma1991} indicating $g=\alpha$.  This is
so far the only known exactly solvable model obeying HES in 2D and it
has a flat dispersion. Therefore, the clarification of the relation
between HES and anyons' braid statistics appears to be a step of great
interest for providing a class of planar models obeying HES with
non-trivial(non-flat) dispersion.

In this article we show that the HES exists quite generically in 2D
systems as long as there is a Hall response.  A general relation between
$g$ and $\alpha$ is established \emph{non-perturbatively} by examining
the total anyon number in the ground state.  Note that determining the
energy distribution function $n_H(\epsilon,g)$ requires a Fock structure
in terms of single anyon states, not found in anyonic systems where the
$N$-body wave function is a multi-valued section depending non-trivially
on the variables of all particles
\cite{Frohlich1989,Schrader1993,*Mund1993}. On the other hand, the total
anyon number is easy to calculate. In fact, according to the Haldane's
definition of the statistical interaction $g$, one immediately obtains
$d(N)-d(0)=(g-1)N$. If the entire band is filled, $N$ reaches its
maximum denoted by $N(g)$, and $d(N)=0$. Note that $d(0)$ is also the
maximum of fermion filling number $N(0)$, hence we find the integral
form of Haldane's statistical interaction
\begin{align}
\label{eq:1}
  \frac{N(g)}{N(0)} = \frac{1}{1-g} \quad \text{or} \quad N(g)-N(0)=\frac{g}{1-g} N(0),
\end{align}
which is consistent with Haldane's energy distribution function
$n_H(\epsilon,g)$, being more general, valid also for Haldane liquids
\cite{Iguchi1998}.  The above description can be easily generalized to
the case of multiple species with \emph{mutual} HES. Labelling the
species with a latin subindex $i,j,...$, with obvious meaning of
symbols, one finds $d_i(\{N_l\}
)-d_i(\{0\})=\sum_j(g_{ij}-\delta_{ij})N_j$, with $g_{ij}(j \neq i)$
defining the \emph{mutual} HES.

In the following we give two proofs of the relation between anyons'
braid statistics and HES, the first one involving an adiabatic argument
given in Sec.~\ref{sec:adiabatic-argument}, while the second one given
in Sec.~\ref{sec:rpa-calculation} is based upon a specific model in a
random phase approximation(RPA). Sec. \ref{sec:disc-concl} is the
conclusion.

\section{Adiabatic argument} 
\label{sec:adiabatic-argument}
According to Wilczek \cite{Wilczek1982}, one
can map an anyon into a fermion bound to a flux tube by a singular gauge
transformation, so that the multivalueness of the anyons' wave-function,
allowed by the non-trivial topology of the configuration space of 2D
identical particles, is removed in the new fermion system. In this
sense, anyons can be viewed as charged fermions with long range gauge
interaction in which the braid statistics is encoded. The flux binding
can be achieved in the Lagrangian formalism by introducing the
Chern-Simons(CS) term. The resulting Lagrangian in the Minkowski
space-time reads
\begin{align}
\label{eq:2}
\mathcal{L}=\mathcal{L}_{M}-a_{\mu}J^{\mu}
+\frac{1}{4\pi\alpha}\epsilon^{\mu\nu\lambda}a_{\mu}
\partial_{\nu} a_{\lambda},
\end{align}
where $J_{\mu}$ is the current and $a_{\mu}$ is the statistical CS gauge
field. By ``charge'' we mean the statistical charge coupled to
$a_{\mu}$'s field, which is taken as unit here.  At present, the precise
form of the fermion Lagrangian $\mathcal{L}_M$ is not important, and it
is only assumed to yield the Hall conductance $\sigma_h$. We could take
$\mathcal{L}_M$ as describing the conventional quantum Hall system, the
magnetized surface state of a topological insulator, or other quantized
anomalous Hall insulators without magnetic
field\cite{Haldane1988a,Qi2006a,Yu2010,Chang2013,Regnault2014,Dobardzic2014}.

By integrating out the $a^0$ field, one obtains the following constraint on
the charge density and the statistical flux
\begin{align}
\label{eq:3}
\vec{\nabla}\times \vec{a} =- 2\pi\alpha J^0.
\end{align}
The Chern-Simons parameter $\alpha$ characterizes the braid statistics,
which measures the Aharonov-Bohm phase when one fermion circles around
another or equivalently two fermions are interchanged.

\begin{figure}[htbp]
\centerline{\includegraphics[width=7.0cm]{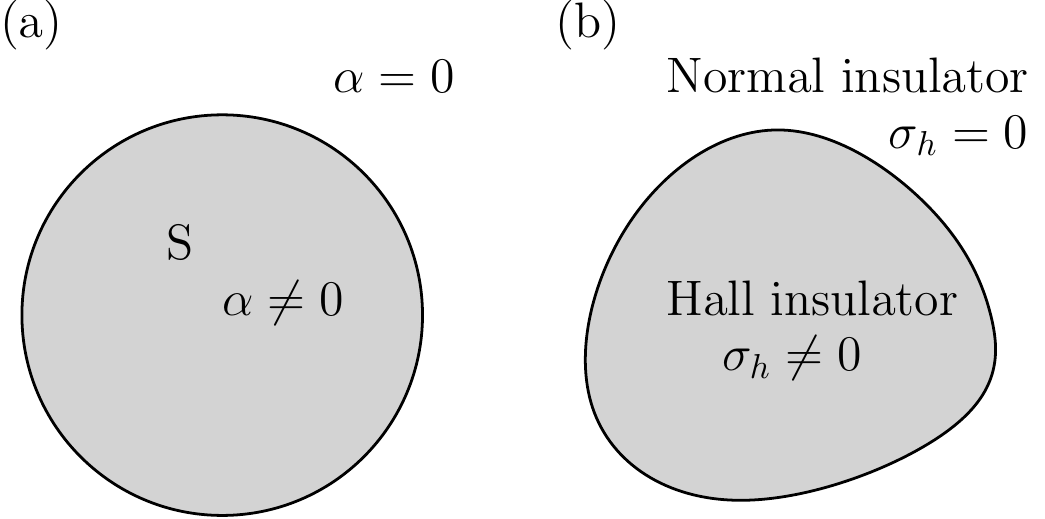}}
\caption[]{\label{fig:1} (a) The Chern-Simons coupling parameter
  $\alpha$ varies adiabatically with time inside the shadowed region
  $S$, being zero outside. 
(b) A Hall insulator connected
  with a normal insulator.}
\end{figure}

To illustrate the relation between the braid statistics and the HES, we
isolate a thermodynamically large region $S$ as shown in
Fig.~\ref{fig:1}a from the rest. Inside $S$ we turn on the statistical
parameter $\alpha$ adiabatically, generating the statistical flux
$\Phi\equiv\int_{\partial S}\vec{a}\cdot d\vec{x}$ through the region
$S$. This in turn induces a current $J^i=-\sigma_h\epsilon^{il}E^{l}$
with $E^l=-\partial_ta^l$ for a \emph{Hall insulator} in the scaling
limit. Then, according to the continuity equation and
  Eq.~\eqref{eq:3} we have
\begin{align}
\label{eq:4}
\partial_t N=- \int_{S} d^2\vec{x}\partial_iJ^i
=2\pi\sigma_h\partial_t(\alpha N),
\end{align}
where $N\equiv \int_Sd^2\vec{x} J^0$ is the total particle number in
$S$.  Eq.~\eqref{eq:4} implies that $N(1-2\pi\sigma_h\alpha)$ is time
independent, therefore the particle number for a general value of
$\alpha$ is given by $N(\alpha)/N(0) =1/(1-2\pi\sigma_h\alpha)$
in the scaling limit.  Comparing with Eq.~\eqref{eq:1},
we obtain the condition
\begin{align}
\label{eq:5}
g=2\pi\sigma_h\alpha,
\end{align}
between the statistical parameters $g$ for HES and $\alpha$ for the
braid statistics. As an example, let's revisit the non-relativistic
anyons in a strong magnetic field. If we take only the LLLs into
account, $\sigma_h=1/(2\pi)$, one finds $g=\alpha$ according to
Eq.~\eqref{eq:5}, which recovers the result given in
Refs.\onlinecite{deVeigy1994,Ma1991}.

In this adiabatic argument, it is crucial for $\mathcal{L}_M$ to
describe a Hall insulator, while a normal insulator with $\sigma_h=0$
does not respond to the Chern-Simons flux in the required way.
Therefore, anyons in a normal insulator can not obey HES.  This proof
can be generalized to \emph{mutual} statistics obeyed by multiple
species of particles labelled by a latin subscript $i$ as after
Eq.~\eqref{eq:1}. Let's consider anyons with statistical charges $q_i$
and Hall conductivity $\sigma_{h,i}$ coupled to a common Chern-Simon
field as in Eq.~\eqref{eq:2}, then the mutual exchange statistics
between species $i$ and $j$ is given by $\alpha_{ij}\equiv \alpha
q_iq_j$. The argument above generalizes Eq.~\eqref{eq:5} to $g_{ij}= 2
\pi \sigma_{h,i} \alpha_{ij} $, thus relating also a mutual
HES\cite{Wu1994} to anyons' braid statistics. The details on the mutual
statistics are given in the appendix.

Next, by invoking the edge states, inevitable in a quantum Hall system
with boundaries, we give another proof of Eq.~\eqref{eq:5} admitting
some generalization beyond the insulating case. In fact, even in
Eq.~\eqref{eq:2} a boundary term would be required to ensure gauge
invariance (see e.g. Ref.\onlinecite{frohlich1995b}), but it has not
been written explicitly because it is irrelevant in the above adiabatic
argument, based only on the continuity equations in the bulk.

\section{RPA calculation} 
\label{sec:rpa-calculation}
We just showed that the Hall response is crucial
for anyons obeying HES. In practice, a Hall insulator can be realized
with or without magnetic field. Since anyons in a strong magnetic field
have been studied before
\cite{Johnson1990,Ma1991,Dunne1991,deVeigy1994}, in this article we
consider the quantum anomalous Hall(QAH) system without Landau levels,
which has been proposed theoretically\cite{Haldane1988a,Yu2010} and
realized in recent experiments\cite{Chang2013}.  To be specific we take
the fermion Lagrangian $\mathcal{L}_M$ as a 2D massive Dirac fermion
with $\mathcal{N}_f$ flavors, which can be written conveniently in the
Euclidean space-time as
\begin{align}
\label{eq:6}
\mathcal{L}_{M} =
\sum_{s=1}^{\mathcal{N}_f}\bar{\psi}_s(\frac{1}{i}\dslash+i\gamma^0\mu+m_s)\psi_s
\end{align}
where $\gamma^{\mu}=\{\sigma_3,i\sigma_1,i\sigma_2\}$,
$\dslash\equiv\partial_{\mu}\gamma^{\mu}$,
$\bar{\psi}=\psi^{\dagger}\gamma^0$, and $m_s$ is the fermion mass of
flavor $s$. In realistic systems, there might be impurities and
conventional Coulomb interactions between fermions, but for simplicity
we ignore them here.  In the following we present a perturbation
calculation of the total anyon number in the ground state.

The chemical potential in $\mathcal{L}_M$ breaks the Lorentz invariance
down to the spatial rotation invariance. Combined with the gauge
invariance of Eq.~\eqref{eq:2}, one finds that the fermion polarization bubble
$\Pi^{(2)}_{\mu\nu}(q|\mu)$ takes the following form\cite{Freedman1977a}
\begin{align}
\label{eq:7}
\Pi^{(2)}_{\mu\nu}(q|\mu)=&(q^2\delta_{\mu\nu}-q_{\mu}q_{\nu})h_1(q|\mu)
\nonumber\\
&+\delta_{\mu k}(\vec{q}^2\delta^{kl}-q^kq^l)\delta_{l\nu} h_2(q|\mu) \nonumber\\
&+\epsilon^{\mu\nu\lambda}q_{\lambda}h_3(q|\mu),
\end{align}
where we denote the space-time indices with Greek symbols, while Latin
characters denote the spatial coordinates.  The three $h_i$'s functions
can be calculated straightforwardly. The first two terms are the
conventional ones of the three-dimensional relativistic fermions with
finite density.  It is the antisymmetric $h_3$-term that is peculiar and
the relevant one for 2D fractional exclusion statistics.

Without chemical potential or for $|\mu|\le\min(|m_s|)$, one finds
$\lim_{q\rightarrow0}h_3=
\mathcal{N}_f/(4\pi)\sum_{s=1}^{\mathcal{N}_f}\sgn(m_s)$
\cite{Redlich1984a,Redlich1984b,Dunne1999,Deser2000}, which provides the
quantized Hall conductance $\sigma_h$ of the system. According to the
Nielsen-Ninomiya theorem\cite{nielsen1981,*nielsen1981a}, the flavor
number $\mathcal{N}_f$ should be even except for some cases with topological
reasons. If the signs of $m_s$'s annihilate in pairs, $\mathcal{L}_{M}$
describes a normal insulator with $\sigma_h=0$. Once these signs do not
cancel completely, we obtain a Hall insulator with integer Hall
conductance.  As an example we note that the Haldane's QAH
model\cite{Haldane1988a} has its low energy physics described by a
$2$-flavor Dirac fermion with the same mass term.  For simplicity, we
assume all  masses $m_s$ have the same positive value $m>0$, then
$\sigma_h=\mathcal{N}_f/(4\pi)$. Unlike other polarization terms($h_1$
and $h_2$), this induced Chern-Simons term is a topological effect which
is independent of the specific characteristics of the model and
non-perturbative in nature. Indeed it depends only on the flavor number
$\mathcal{N}_f$. We then adopt the random phase approximation by taking
only this topological term into account and obtain the following
propagator of the gauge field,
\begin{align}
\label{eq:8}
  \tilde{D}_{\mu\nu}(x,x') =
  i2\pi\tilde{\alpha}\partial^{-2}\epsilon^{\mu\nu\lambda} \partial_{\lambda}
  \delta^{(3)}(x-x'),
\end{align}
where $\tilde{\alpha}=\alpha/(1-2\pi \sigma_h\alpha)$ and the Landau
gauge is assumed for convenience.

To calculate the anyon number, we first evaluate the free energy
$\mathcal{F}(\mu,\alpha)$ to the lowest perturbation order with all 
gauge boson lines being replaced by $\tilde{D}_{\mu\nu}$ of
Eq.~\eqref{eq:8}, which includes the direct and exchange terms $\mathcal{F}_{d}$
and $\mathcal{F}_e$
\begin{align}
\label{eq:9}
\mathcal{F}_{d}
=&-\frac{1}{2\beta} \int d^3x d^3x'
J^{\mu}(x) \tilde{D}_{\mu\nu}(x-x') J^{\nu}(x') ,  \\
\mathcal{F}_e =& \frac{\mathcal{N}_f}{2\beta}
\int  d^3xd^3x'  \tilde{D}_{\mu\nu}(x-x') \nonumber\\
\label{eq:10}
&\tr[S_0(x'-x|\mu)\gamma^{\mu}S_0(x-x'|\mu)\gamma^{\nu}],
\end{align}
where $\beta$ is the inverse temperature and $S_0(x|\mu)$ is the free
fermion propagator of a single flavor.  Although it is almost the
simplest approximation, it can reproduce the \emph{non-perturbative}
results for the anyon insulator given earlier, since it already
incorporates the important topological effect in the propagator
$\tilde{D}_{\mu\nu}$.

We first consider the insulating case with $\mu=0$. To investigate the
anyon number in the filled band(or vacuum), we require that the
statistical charges of particles in the valence band are not screened at
all, so that the vacuum expectation value of anyon density
$J^0(x)\ne0$. However $J^i(x)$ still vanishes if one calculates the
fermion loop directly since a filled band usually can not carry any
persistent current. Note that the RPA propagator $\tilde{D}_{\mu\nu}$ is
antisymmetric and one might conclude $\mathcal{F}_d=0$ at first
glance. This is true for the normal insulators, but it is not correct
for the Hall insulator where a persistent chiral current does exist at
the sample edges\cite{wen1992,frohlich1995b}, as required by gauge
invariance, though it vanishes in the bulk. Since $\tilde{D}_{\mu\nu}$
contains an infrared singularity, one can kill two birds with one stone
by putting the system in a finite area $\Omega$ in contact with normal
insulator as shown in Fig.~\ref{fig:1}b.  For the static state, the
current densities are time independent and, if the characteristic
function of $\Omega$ is denoted by $f_{\Omega}(\vec{x})$, the chiral
current density can be written as
$\vec{J}(\vec{x})=I(\mu)(\hat{z}\times\vec{\nabla}f_{\Omega})$ with $I$
the edge current. Then the direct term can be written as
\begin{align}
\label{eq:11}
\mathcal{F}_d(\mu,\alpha)=& -i2\pi\tilde{\alpha}\int d^2\vec{x}
J^0(\vec{x}) \vec{\nabla}^{-2}\epsilon^{kj}\partial_jJ^k(\vec{x}) \nonumber\\
=&-2\pi\tilde{\alpha} I(\mu)N(\mu,0),
\end{align}
which leads to the modification of anyon number
\begin{align}
\label{eq:12}
\Delta N_{d}(\mu,\alpha) =& 2\pi\tilde{\alpha} \sigma_h
N(\mu,0)
 + 2\pi\tilde{\alpha} I(\mu) \frac{\partial N}{\partial\mu}.
\end{align}
In the presence of the Chern-Simons term, the chiral current also
induces an electric field perpendicular to the boundary, which is
proportional to $2\pi\alpha I(\mu)$ and leads to an additional inner
pressure. This is the origin of the second term in the r.h.s. of
Eq.~\eqref{eq:12}. Indeed, $\partial N/\partial \mu$ is proportional to
the compressibility $\kappa_T$ for a fixed area. For the insulating
state, where $\kappa_T$ is zero, we then recover the results given earlier and we can extend it to the more general case of
incompressible, but not necessarily insulating systems. It is also noted that in this case, there is no
contribution to the particle number from the exchange term
$\mathcal{F}_e$ and from other possible interactions, since they do not
depend on $\mu$ as long as the bulk gap is not spoiled.

We now consider a compressible anyon gas with $\mu>m$, where the bulk
excitation gap vanishes. In this case, we are only interested in the
anyon gas in the conduction band, and assume the valence band is
screened by a static background with opposite charges. The finite Fermi
sea modifies the polarization bubble Eq.~\eqref{eq:7}
\cite{Frohlich1995,Ye2013}. The Hall conductance $\sigma_h$ is no longer
quantized and the compressibility $\kappa_{T}$ is also nonzero
accordingly. In this case not only the second term in the r.h.s. of
Eq.~\eqref{eq:12} but also the exchange term $\mathcal{F}_e$
contributes.

The calculation of $\mathcal{F}_e$ follows the standard procedure for
the QED with finite density given in
Refs.\onlinecite{Freedman1977a,Morley1978}
\begin{align}
\label{eq:13}
\mathcal{F}_e(\mu,\alpha) = -\frac{\tilde{\alpha} m \mathcal{N}_f \Omega}{4\pi}
(\mu-m)^2.
\end{align}
Differentiating $\mathcal{F}_e$ with respect to $\mu$ leads to the
modification of particle number from the exchange term,
\begin{align}
\label{eq:14}
  \Delta N_e(\mu,\alpha) =\frac{2\tilde{\alpha}m}{\mu+m} N(\mu,0)
\end{align}
with $N(\mu,0)\equiv \mathcal{N}_f \Omega(\mu^2-m^2)/(4\pi)$. The total
particle number change is the sum of Eq.~\eqref{eq:12} and
Eq.~\eqref{eq:14}. 

Finally, we consider a charge neutral situation in which all statistical
charges in both valence and conduction bands are screened completely, so
that all fermion loops vanish and only the exchange term
survives. Further assuming $\mu-m\ll m$, we obtain $N(\mu,\alpha)\approx
( 1+ \tilde{\alpha}) N(\mu,0)$.  If $\mathcal{N}_f=2$, we obtain
$g=\alpha$, a result similar to the insulating case is obtained. The
relation between HES and the exchange diagram has also been discussed in
Ref.\onlinecite{Chen1995} where they set $\mathcal{N}_f=1$ and did not
consider the RPA.

\section{Discussion and Conclusion} 
\label{sec:disc-concl}
We have discussed the exclusion statistics of anyons in a generic
quantum Hall system.  Instead of analyzing the distribution of single
particle states requiring Fock structure, we study the total particle
number in the many body ground state in a finite volume, which has been
shown to characterize the crucial aspect of the statistics. Our result
proves that the total particle number of the ground state of anyons
satisfies HES if the bulk excitations are fully gapped or more generally
for an incompressible system.  The crucial role played by edge currents
for this result is clarified. Anyons in normal insulators without Hall
conductance and edge states do not satisfy the HES.

The compressible anyon gas does not show the HES as clean as the
incompressible liquid. Actually, it is noticed that the
  braiding statistics is not well defined at all in the 2D compressible
  superfluids\cite{Haldane1985a}. Anyhow, for compressible systems 
if the parameter $\alpha$ is well defined the relation between $g$ and
$\alpha$ is much more complicated than Eq.~\eqref{eq:5}, involving also
the system-dependent parameters like fermion mass. Nonetheless at fixed
chemical potential the particle number in finite volume strongly depends
on the statistical parameter $\alpha$, and it even diverges at
$2\pi\sigma_h\alpha=1$(see e.g. Eq.~\eqref{eq:12}) as a Bose
liquid. Since anyons can be viewed as fermions with gauge interaction,
this indicates the violation of the original Luttinger theorem for
fermions\cite{Luttinger1960} by the Chern-Simons interaction with
non-integer $\alpha$.

Our approach provides specific microscopic models satisying HES with
non-trivial dispersion, it allows easily judging whether a system can
obey HES or not, and points a direction for searching such an exotic
statistics in real materials.  We hope that our work may shed light on a
large class of models in 2D, as the relation with HES did in 1D. In
particular, it allows a cleaner treatment of the semionic ($\alpha=1/2)$
holons in the 2D $t-J$ model, relevant to the cuprate high Tc
superconductors\cite{Marchetti1998,Marchetti2011,Marchetti2012}, in
analogy with the semionic holons of the 1D $t-J$ model
\cite{Ha1994,Marchetti1996}.

\section{Acknowledgements} F.Y. is supported by NSFC
11374135. P.A.M. acknowledges the partial support from the Ministero
Istruzione Universit\'a Ricerca (PRIN Project "Collective Quantum
Phenomena: From Strongly-Correlated Systems to Quantum Simulators").

\appendix*
\section{Adiabatic argument for mutual statistics}
In this appendix we give the details of the adiabatic argument for the
mutual statistics. We first consider $s$ kinds of particles satisfying
the mutual Haldane's exclusion statistics(HES)\cite{Haldane1991a,Wu1994}
characterized by the following equation
\begin{align}
\label{eq:15}
\Delta d_i = \sum_j (g_{ij}-\delta_{ij})\Delta N_j
\end{align}
where $d_i$ and $N_i$ are the dimensionality of the new adding particle
and the total particle number of kind-$i$, respectively. $g_{ij}$ for
$i\ne j$ is the \emph{mutual HES} \cite{Wu1994}. 

Integrating Eq.~\eqref{eq:15}, one finds the dimensionality $d_i$ as a
function of the particle numbers $\{N_j\}$, 
\begin{align}
\label{eq:16}
d_i(\{N_j\}) =& \sum_j(g_{ij}-\delta_{ij})N_j+N_{i,0} 
\end{align}
where $N_{i,0}\equiv d_i(0)$ is the dimension of the available Hilbert
space for kind-$i$ particle when there is no particles, which is also
the maximal filling number for fermions when $g_{ij}=0$(namely, no
exotic HES).  If the system is insulating, the dimensionality
$d_i(\{N_j\})=0$, and the particle number $N_{j,\text{g}}$ can be solved
from Eq.~\eqref{eq:16} with the following form
\begin{align}
\label{eq:17}
N_{i,0} = \sum_j(\delta_{ij}-g_{ij}) N_{j,\text{g}}.
\end{align}

Next, let's consider the \emph{braid} statistics.  Suppose we have a
model involving $s$ kinds of anyons which can be described by fermions
coupled to a common Chern-Simons field with the following Lagrangian
\begin{align}
\label{eq:18}
  \mathcal{L} =
  \mathcal{L}_{M}[\psi_j,\psi_j^{\dagger}]
  -\sum_{j=1}^s q_j J^{(j)}_{\mu}a^{\mu} + \frac{1}{4\pi\alpha}
  \epsilon^{\mu\nu\lambda} a_{\mu}\partial_{\nu}a_{\lambda}
\end{align}
where $J^{(j)}$ is the current of kind-$j$ particle, and $q_j$ is the
corresponding statistical charge. As in
Sec.~\ref{sec:adiabatic-argument}, we also assume $\mathcal{L}_M$
providing a nontrivial Hall response $\sigma_{h,j}q_j^2$ for kind-$j$
particle. Each particle is now bound with flux $\phi_{j}$ depending on its
statistical charge $q_j$
\begin{align}
\label{eq:19}
\phi_j = -2\pi\alpha q_j.
\end{align}
Following the adiabatic argument given in
Sec.~\ref{sec:adiabatic-argument}, we isolate a thermodynamically large
area $S$, through which the total flux reads
\begin{align}
\label{eq:20}
\Phi =\sum_{j=1}^sN_j\phi_j= -2\pi\alpha \sum_{j=1}^{s}q_jN_j.
\end{align}
As we turn on $\alpha$ adiabatically, according to the continuity equation, we
obtain
\begin{align}
\label{eq:21}
  \partial_t(q_iN_i) =& -(\sigma_{h,i}q_i^2)\partial_t\Phi =
                        2\pi(\sigma_{h,i}q_i^2)\partial_t \left( \alpha \sum_{j=1}^{s}q_jN_j \right).
\end{align}
For convenince, we can introduce the \emph{exchange matrix}
$\alpha_{ij}\equiv \alpha q_iq_j$, which reflects the anyons' mutual
\emph{braid} statistics between kind-$i$ and kind-$j$ particles.
Then, Eq.~\eqref{eq:21} can be rewritten as
\begin{align}
\label{eq:22}
\partial_t(N_i-2\pi \sigma_{h,i}\sum_{j}\alpha_{ij} N_j)=0,
\end{align}
which implies the term in the bracket is time-independent when $\alpha$
is changed from $0$ to a finite value adiabatically, therefore
\begin{align}
\label{eq:23}
N_i(\alpha)-2\pi \sigma_{h,i}\sum_j\alpha_{ij}N_{j}(\alpha) = N_i(0),
\end{align}
where $N_i(0)$ is the total fermion number of kind-$i$ when $\alpha=0$. 

Identifying $N_i(\alpha)$($N_i(0)$) in Eq.~(\ref{eq:23}) with
$N_{i,g}$($N_{i,0}$) in Eq.~(\ref{eq:17}), we establish a relation
between mutual HES $g_{ij}$ and the anyons' mutual \emph{braid}
statistics $\alpha_{ij}$,
\begin{align}
\label{eq:24}
g_{ij}=2\pi \sigma_{h,i}\alpha_{ij}.
\end{align}

\end{document}